\documentclass{elsart}
\usepackage{graphics}

\usepackage{amssymb}

\begin{document}

\begin{frontmatter}



\title{On non-Fermi liquid quantum critical points in heavy fermion metals}


\author[iisc,mit]{T. Senthil}

\address[iisc]{Center for Condensed Matter Theory, Indian Institute of Science,
Bangalore-560012, India}
\address[mit]{Department of Physics, Massachusetts Institute of
Technology, Cambridge, Massachusetts 02139}

\begin{abstract}
Heavy electron metals on the verge of a quantum phase transition to magnetism show a number of unusual non-fermi liquid properties which are poorly understood. This article discusses in
a general way various theoretical aspects of this phase transition with an eye toward understanding the non-fermi liquid phenomena.  We suggest that the non-Fermi liquid quantum critical state may have a sharp Fermi surface with power law quasiparticles but with a volume not set by the usual Luttinger rule. We also discuss the possibility that the electronic structure change associated with the possible Fermi surface reconstruction may diverge at a different time/length scale from that associated with magnetic phenomena. 
\end{abstract}

\end{frontmatter}

\section{Introduction}
\label{intro}

In recent years, a number of experiments have studied the quantum
phase transition between a non-magnetic Fermi liquid metal and an
antiferromagnetically ordered metal in the heavy fermion materials\cite{hfexp}.
In the best studied cases, a host of strange phenomena have been
discovered in the vicinity of the transition. Most remarkably the
transition has been found to be accompanied by a rather severe
breakdown of Fermi liquid theory. A proper theoretical description
of this phase transition and the observed non-Fermi liquid physics
is currently lacking and is the subject of much theoretical
discussion\cite{colschonat}.

Microscopically the heavy fermion materials may be modelled as a
lattice of localized magnetic moments that are coupled to an
itinerant band of conduction electrons through spin exchange. A
crucial question is the ultimate fate of the localized moments in
the quantum ground state of the coupled system. In the fermi
liquid state the local moments are absorbed into the Fermi sea by
a lattice analog of the Kondo effect. The resulting Fermi surface
has a volume consistent with Luttingers theorem only if the local
moments are included in a count of the total electron density.
This is known as the `large' Fermi surface. Further the
quasiparticles near the Fermi surface typically have large
effective masses - this is the origin of the term `heavy fermi
liquid'.

In the magnetically ordered states on the other hand the local
moments are frozen in time in specific directions. An important
though often overlooked point is that there may actually be two
distinct kinds of magnetic metals. Crudely speaking in one kind
the magnetism may be viewed as arising from imperfectly Kondo
screened local moments, essentially as a spin density wave formed
out of the parent heavy Fermi liquid. This state will be dubbed
SDW. The other kind of magnetic state arises when the moments
order directly due to RKKY exchange interactions and are not part
of the fermi surface of the metal. As in Ref. \cite{svsrev} we
will refer to this state as the local moment magnetic metal (LMM).
The distinction between the two magnetic metals can be sharp in the
sense that the two are not connected smoothly without any
intervening phase transitions.

The purpose of the present paper is to discuss in a general way
several aspects of the quantum phase transition associated with
the onset of magnetism in the heavy Fermi liquid with a focus on
understanding the non-Fermi liquid phenomena observed in
experiments. We begin by elaborating on the distinction between
the two kinds of magnetic metals mentioned above. We then discuss
the possibility that the strong breakdown of Fermi liquid theory
observed in experiments is associated with the transition to the
LMM state. In contrast the transition to the SDW state is expected
to be described by the well-developed Moriya-Hertz-Millis theory\cite{Hertz}
which provides for only weak deviations from Fermi liquid theory.

A theory of the quantum phase transition between the heavy Fermi
liquid and LMM states does not exist at present. Here we will
first examine in a rather general manner some properties that such
a phase transition will have if it is second order. We will then
examine some possible ways in which such a second order transition
might actually occur. We will suggest various experiments that
could potentially pave the way for a theoretical understanding of
the non-Fermi liquid physics.

\section{Magnetic metals in Kondo lattices}
\label{mmkl}

The Kondo lattice model provides a useful framework to think about general issues in heavy fermion physics:
\begin{eqnarray}
H & = & H_t + H_K + H_J \\
H_t & = & \sum_{k}\epsilon_k c^{\dagger}_k c_k \\
H_K & = & \frac{J_K}{2} \sum_r \vec S_r. c^{\dagger}_r \vec \sigma
c_r \\
H_J & = & J\sum_{<rr'>} \vec S_r. \vec S_{r'}
\end{eqnarray}

The first term represents the kinetic energy of conduction
electrons $c_k$ on some lattice. The second term is a Kondo spin
exchange between conduction electrons and localized spin-$1/2$
moments $\vec S_r$ on the sites $r$ of the lattice. It is
conceptually convenient to allow for a Heisenberg exchange between
the local moments (the last term in the Hamiltonian).

\begin{figure}
\includegraphics{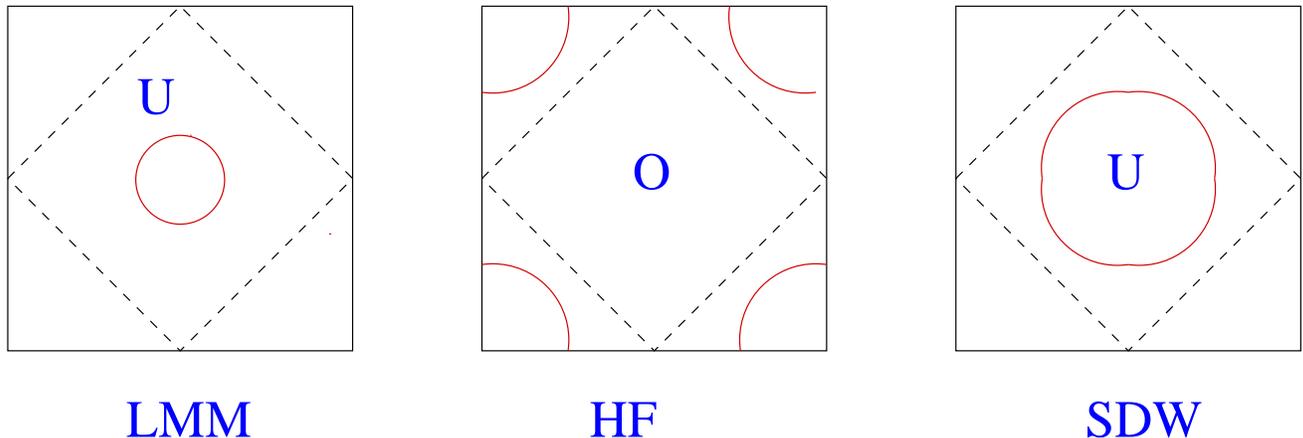}
 \caption{Fermi surface (red curves) of the Local Moment Magnetic (LMM) metal, the heavy fermi liquid (HF) and the spin density wave (SDW) metal for a two dimensional square lattice. O denotes the occupied region and U the unoccupied region.  For the SDW and LMM the Fermi surfaces is drawn in the magnetic Brillouin zone (dashed line) while for the HF it is shown in the full Brillouin zone. Note that both states satisfy Luttinger theorem.} \label{fsm}\end{figure}

Let us first describe the LMM metal. Imagine that the Heisenberg
exchange dominates over the Kondo exchange in the Hamiltonian. In
that situation, the local moments will predominantly talk to each
other rather than to the conduction electrons. With unfrustrated
Heisenberg interactions, the local moments will then order
magnetically. If we further assume that the conduction band does
not interact the magnetic Brillouin zone, then the $c$-electrons
will essentially be unaffected by the magnetic ordering. Thus the
local moments and $c$-electrons will more or less stay decoupled
in this state. The Fermi surface of this metal will basically be
determined by the $c$-electron band structure (see Fig. \ref{fsm}).

Now consider the other kind of magnetic metal - the SDW state. To
understand it we first need to understand its parent state, the
heavy Fermi liquid. To that end let us first ignore the Heisenberg
exchange term. The Kondo process responsible for the occurence of
the Fermi liquid is conveniently described in a mean field theory
which represents the local moments in terms of fermionic degrees
of freedom $f_{r\alpha}$ ($\alpha = 1,2$):
\begin{equation}
\vec S_r = \frac{1}{2}f^{\dagger}_r \vec \sigma f_r
\end{equation}
together with the constraint $f^{\dagger}_rf_r = 1$ at each site.
The mean field approximation consists of choosing an appropriate
self-consistently determined quadratic Hamiltonian in terms of the
$c$ and $f$ operators. The Kondo process is described in terms of
a `hybridization' between the $c$ and $f$ bands. In the heavy
Fermi liquid, the hybridization amplitude is non-zero, and leads
to quasiparticles whose band structure is modified from that of
the bare conduction electrons. The Fermi surface is large in the
sense that its volume counts both the $c$ and $f$ particles (see
Fig. \ref{fsm}).

Now consider the effect of increasing $J_H$ on this heavy Fermi
liquid. It is conceivable that at some point it becomes favorable
to form a spin density wave state out of the renormalized
quasiparticles of the heavy Fermi liquid (while the Kondo
hybridization stays non-zero). Is this state smoothly connected to
the LMM described above? To answer this let us examine the Fermi
surface of this SDW state in the magnetic Brillouin zone (assuming
the same magnetic ordering pattern as in the LMM). This is shown
in Fig \ref{fsm}. Comparing with the Fermi surface of the LMM state we see
that the two Fermi surfaces have different orientation. Thus at
least in this model these two kinds of magnetic metals {\em
cannot} be smoothly connected to each other. A Lifshitz transition
associated with reconstruction of the Fermi surface must separate
the two phases. Note though that both states satisfy Luttingers theorem which relates the total Fermi surface volume 
{\em modulo the Brillouin zone volume} to the total density of electrons. 

Though the preceding discussion focused on the situation where the
large Fermi surface did not intersect the magnetic Brillouin zone
a similar distinction will in general hold in that situation as
well. In complex real systems with many partially filled bands and
(possibly) incommensurate magnetic order it is harder to clearly
demonstrate a distinction between the two phases based on the
orientation/topology of the Fermi surface. However it seems rather
likely that the distinction will survive. Finding a useful
criterion to distinguish these two magnetic phases in such a
general setting is an important open theoretical question.

\section{Magnetic phase transition: generalities}
\label{qptgen}

\begin{figure}
\includegraphics{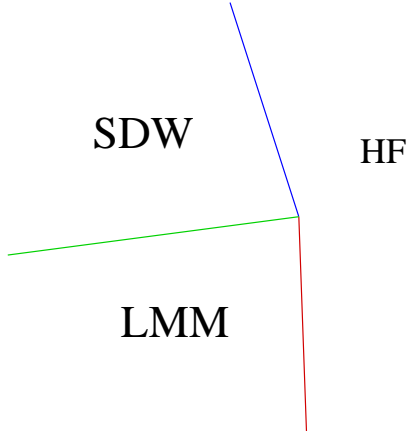}
 \caption{Schematic phase diagam showing the heavy fermi liquid (HF), spin density wave (SDW)
and Local Moment Magnetic metal (LMM) phases. The transition between HF and SDW is expected 
to have a description within the Moriya-Hertz-Millis dsecription unlike that between the HF and LMM phases.} \label{hfpd}\end{figure}

What may we say about the quantum phase transition associated with
the onset of magnetic order from the Fermi liquid? A priori we
should expect that the universality class of the transition
depends on which of the two distinct kinds of magnetic phases the
transition is to (see Fig. \ref{hfpd}. Considerable effort has gone into examining the
transition to the spin density wave metal from the Fermi liquid\cite{Hertz}.
This theory (known as the Moriya-Hertz-Millis theory) focuses on
the long wavelength low energy fluctuations of the natural
magnetic order parameter in a metallic environment and makes
specific predictions for a number of physical properties near the
quantum phase transition. However these predictions seem
inconsistent with the strong non-Fermi liquid physics seen in some
heavy fermion metals (such as $CeCu_{6-x}Au_x$ or $YbRh_2Si_2$)
near the magnetic ordering transition. Thus at least in these
materials it is worthwhile to explore the possibility that the
observed transition is to the LMM phase. As the Fermi surface of
the LMM is rather different from that of the heavy Fermi liquid,
such a transition is associated with a substantial reconstruction
of the Fermi surface. This is over and above the onset of magnetic
order.

Empirically the hypothesis that non-Fermi liquid physics is
associated with a transition to the LMM may in principle be
checked in some ways. First the Hall
coefficient is expected to jump on passing through the transition\cite{cpsr}.
Supporting evidence has appeared in recent measurements\cite{Silke} on
$YbRh_2Si_2$. It is important to point out though that as a matter
of principle confirming/ruling out this hypothesis does not
require detailed measurements of the Fermi surface structure near
the quantum critical point. Rather one can go far away from the
transition into either the magnetic or Fermi liquid phases and
obtain information about their respective Fermi surfaces. Now upon
tuning toward the quantum critical point if there are no other
phase transitions then the Fermi surface data can be directly
compared to see if major reconstruction has occured.

The properties of a phase transition between the Fermi liquid
and the LMM state are rather poorly understood. Can it even be
second order? Two rather distinct phenomena (magnetic ordering and
Fermi surface reconstruction) are required to occur at the same
point as we tune some parameter across the transition. Thus
naively it might appear that a direct second order transition
might occur only if more than one parameter is fine-tuned (as at a
multicritical point). Interestingly recent theoretical work on
simpler problems has raised the possibility that this naive
expectation might be incorrect\cite{deccp}. If such a second order transition
were to exist, it must clearly have rather unusual properties. How
do we describe such a transition? Will it reproduce the observed
non-Fermi liquid physics in experiments? We do not currently know
the answers to these questions. In the rest of the paper we will
discuss some ideas on how the answers might eventually work out
with a view to motivating future experimental and theoretical
work.

\section{Critical non-fermi liquid: ``superlarge" Fermi surface?}
\label{slfs}

The reconstruction of the Fermi surface necessary to go from the
heavy Fermi liquid to the LMM is quite drastic. Even without
considering the added complication of magnetic ordering how can
the transition be second order? One possible way out is the
following. On approaching the quantum critical point from the
Fermi liquid side suppose the quasiparticle residue $Z$ vanishes
everywhere on the large Fermi surface. Then the quasiparticle on
this Fermi surface dies on approaching the quantum critical point.
Moving further out into the LMM side one can then imagine that
there is no quasiparticle-like peak in the electron spectral
function at the location of the large Fermi surface. If the $Z$
vanishes continuously this would be a mechanism for the large
Fermi surface to disappear continuously on going from the Fermi
liquid side to the LMM. Exactly the same process can also be
associated with the continuous disappearance of the Fermi surface
of the LMM metal. Specifically suppose the quasiparticle residue
$Z$ vanishes continuously everywhere on this Fermi surface when
approaching the critical point.

Within this line of thought what is the nature of the electronic
excitations right at the quantum critical point? Consider first
the fate of the quasiparticle peak of the large Fermi surface. For
a second order transition where $Z$ vanishes continuously it seems
reasonable that the $\delta$-function quasiparticle peak is
replaced by a singular power law peak. Thus one might expect that
right at the quantum critical point the large Fermi surface survives
but with a singular power law quasiparticle. Repeating this
reasoning from the LMM side, we are lead to suppose that in
addition a small Fermi surface {\em also} survives with its own
power law quasiparticle.

These considerations provide a rather interesting picture of the
electronic excitations of the critical ground state. It has two
identifiable ``Fermi surfaces" - one large and the other small -
but with power law quasiparticle peaks (rather than
$\delta$-function) in the electron spectral function. Clearly this
is a strongly non-Fermi liquid state. The presence of a sharply
defined Fermi surface (through the power law singularity) in this
non-Fermi liquid suggests that it may be viewed as a higher
dimensional analog of the familiar Luttinger liquid state of one
dimensional systems. In two dimensions such a Luttinger liquid
picture was advocated by Anderson\cite{pwa} as a means to describe the
optimally doped cuprates. However in striking contrast to one
dimensional Luttinger liquids or their two dimensional avatars
envisaged by Anderson, the quantum critical metal discussed in
this section {\em does not} satisfy the Luttinger rule on the total
volume of the Fermi surface. Indeed if the number of conduction
electrons per unit cell is $n_c$, the presence of both the small
and large Fermi surfaces implies a total Fermi volume
corresponding to $(1+ n_c) + n_c = 1+ 2n_c$ electrons per unit
cell. We may call this a `super-large' Fermi surface.

One can also conceive of an alternate (tamer) process through
which the Fermi surface reconstruction might take place in a
second order transition. This is motivated by the calculations in
Ref. \cite{ssvprb} for a specific model. Suppose that on the Fermi liquid
side the large Fermi surface actually consists of more that one
sheet. For concreteness consider the case with two sheets - one of
these sheets (denoted the cold sheet) simply evolves into the
Fermi surface of the LMM on going through the transition. Its
quasiparticle peak is retained through out the transition
including right at the critical point. Upon approaching the
transition from the Fermi liquid the volume of this cold sheet
changes continuously to match on to that set by the conduction
electron density alone at the critical point. The other sheet -
denoted the hot sheet - is where all the important action
associated with the critical phenomena happen. This hot sheet
disappears continuously on approaching the critical point from the
Fermi liquid side through a vanishing of its quasiparticle
residue. Further the volume of the hot sheet evolves continuously
so that right at the critical point it only accomodates the local
moments. Again right at the critical point we might expect that
the hot Fermi surface survives but with a power law singularity in
the electron spectral function. The nature of the excitations of
the quantum critical ground state will then be somewhat simple.
There is a cold Fermi surface with volume set by conduction
electron density and with a well-defined quasiparticle peak. In
addition there is a hot Fermi surface with volume set by the local
moment density and with a power law quasiparticle singularity. The
total Fermi surface volume is the same as that of the large Fermi
surface and thus satisfies Luttinger's theorem. This would thus be
a non-Fermi liquid ground state with a sharp Fermi surface
satisfying Luttinger theorem. However the breakdown of the Fermi
liquid occurs in only one sheet and not in all sheets of the Fermi surface.

\section{Two diverging scales}
\label{tds}

As mentioned above at the Fermi liquid-LMM transition two rather
different things seem to happen simultaneously. The magnetic order
appears concomitantly with the electronic structure change
associated with Fermi surface reconstruction. Here we will
distinguish two different ways in which such a second order
transition might work out. We will call these the single scale and
two-scale hypotheses.

Let the characteristic time scale of the nagnetism be denoted
$t_m$ and for the electronic structure by $t_{el}$. For instance
$t_m$ may be taken as the correlation time of the fluctuations of
the magnetic order parameter and can be measured by neutron
scattering experiments. Similarly $t_ {el}$ may be defined  from
the electron spectral function measurable ((at least in principle)
through photoemission experiments. Specifically the electron
spectral function of the Fermi liquid very near the Fermi surface
will crossover from the critical power law form to the
quasiparticle peak at some characteristic energy scale. The
inverse of this energy scale may be taken to define $t_{el}$. If
the transition were second order we might expect that these
characteristic time scales $t_m$ and $t_{el}$  both diverge.

Now typically near many familiar critical points there is only a
single diverging length or time scale. Thus one possibility is
that $t_m, t_{el}$ diverge identically with their ratio remaining
constant. We will call this the single scae hypothesis. However in
view of the rather different physical phenomena that they
characterize it may be surprising to find the same divergence for
both time scales. It is therefore very interesting to contemplate
the possibility that $t_m$ and $t_{el}$ diverge with different
exponents so that their ratio also has singular behavior at the
quantum critical point. We will call this the two-scale
hypothesis. This was first formulated in Ref. \cite{svsrev}. To be precise let
\begin{equation}
t_m \sim |g - g_c|^{-\phi_m}
\end{equation}
where $g$ is the control parameter used to drive the transition
(pressure, magnetic field, etc) and $g_c$ is the critical point.
Similarly
\begin{equation}
t_{el} \sim |g - g_c|^{-\phi_{el}}
\end{equation}
The single scale hypothesis corresponds to $\phi_m = \phi_l$ while
the two-scale hypothesis corresponds to $\phi_m \neq \phi_{el}$.
In the second case we should further distinguish the two cases
$\phi_m > \phi_l$ and $\phi_m < \phi_l$. For concreteness let us
consider the possibility that $\phi_{m}
> \phi_{el}$. In this case the magnetic time scale diverges faster
than the electronic one. The electronic structure changes first
and is followed later by the development of magnetic order. As the
ratio $t_m/t_{el}$ diverges at the quantum critical point the
magnetic phenomena and the electronic structure change are well
separated dynamically, occuring on vastly differing time scales.

The two-scale hypothesis is theoretically appealing for a few
different reasons. First it provides a possible resolution of one
of the key difficulties associated with the transition between the
heavy Fermi liquid and the LMM state. Namely how can a transition
that involves the vanishing of two distinct competing orders
(magnetic in the LMM and `Kondo order' in the Fermi liquid) be
generically second order? Shouldn't the two competing orders be
separated as a function of tuning parameter? The two-scale
hypothesis proposes that the two competing orders are indeed
separated but not as a function of tuning parameter. Rather the
separation is dynamical as a function of scale (time or length
scale).

The two diverging time scales will manifest themselves as two
distinct vanishing energy scales as the quantum critical point is
approached. In turn these will set two distinct vanishing
temperature scales $T_m$ and $T_{el}$ corresponding to magnetic
and Kondo phenomena respectively.

If correct the two-scale hypothesis also has practical benefits
for developing a theory of the quantum transition. The separation
as a function of scale suggests that it may be possible to treat
the two crossovers at $t_{el}, t_m$ separately. Either crossover
then involves only one fluctuating order and is therefore easier
to think about. Let us now consider in a bit more detail the two
separate possibilities with $\phi_m$ greater (less) than
$\phi_{el}$.

\subsection{Kondo-driven: $\phi_m > \phi_{el}$}
\label{kd}
If $\phi_m > \phi_{el}$, then the {\em underlying} primary
transition is associated with the Fermi surface reconstruction.
The magnetism is a secondary low energy complication that happens
once the local moments drop out of the Fermi surface. The
transition may then be regarded as being driven by the loss of
Kondo screening.

What is the nature of the state that obtains between the two
diverging time scales $t_{el}$ and $t_m$? In this intermediate
state the local moments are quenched neither by Kondo screening
nor by static antiferromagnetic or any other form of true long
range order. A reasonable assumption then is that the local
moments have developed singlet bond correlations without settling
down into any particular long range ordered state. Such a state
may be thought of as a resonating valence bond spin liquid formed
out of the local moments. In the absence of Kondo screening the
conduction electrons are essentially decoupled from the local
moments, and fill a {\em small} Fermi surface. Thus this
intermediate time scale state is rather exotic. Such a spin liquid
state coexisting with a small Fermi surface of conduction
electrons was dubbed a `fractionalized Fermi liquid' (FL$^*$) in
Ref. \cite{flstar}. Different kinds of FL$^*$ phases may be distinguished
based on the different kinds of possible spin liquid structures.

In thinking about the possible quantum phase transition with
$\phi_m > \phi_{el}$ theoretically the Fermi surface
reconstruction phenomena may be treated first without reference to
the magnetic ordering. The theoretical problem thus reduces to two
subproblems which are hopefully simpler. First we need to describe
the transition between the heavy Fermi liquid and an appropriate
FL$^*$ phase. Next we need to show how magnetic long range order
appears as a low energy instability of this FL$^*$ phase. Note
that as the ratio of the two time scales $t_m/t_{el}$ also
diverges at the transition, the unstable FL$^*$ phase obtains in a
wide window asymptotically close to the transition.

\subsection{Magnetism driven: $\phi_{el} > \phi_m$}
\label{md}

Next we consider the situation where $\phi_{el} > \phi_m$ so that
$t_{el}$ diverges faster than $t_m$. In this case the primary
transition is the loss of magnetic order. The Kondo screening
develops as a secondary phenomenon once the magnetic long range
order is destroyed. In this situation at intermediate times in the
paramagnetic side, a state with no magnetic order and a small
Fermi surface of conduction electrons is obtained. Presumably the
local moments again have short ranged antiferromagnetic
correlations and may be thought of as forming a spin liquid. This
too is an FL$^*$ state albeit a different one from the one
discussed above. This state must then eventually be unstable (at
scales $t_{el}$) towards the development of the large Fermi
surface heavy Fermi liquid for this scenario to work.

\section{Towards a theory}
\label{tat}

It has thus far not been possible to develop the thinking outlined
in previous sections into a serious theory of the non-fermi liquid
heavy electron quantum critical points. However substantial
progress has been achieved in understanding various theoretical
issues that may pave the way for future work in this direction. In
the following sections we will quickly review these results.

\section{The two-scale hypothesis}
From a renormalization group standpoint the two-scale hypothesis 
would obtain most naturally where the ordering in one of the two phases is due to a perturbation that is dangerously irrelevant at the critical fixed point. 
Below we discuss the theoretical evidence for this phenomenon at quantum critical points.

\subsection{Deconfined quantum criticality}
 The naive intuition
against a direct second order transition between the HF and LMM states is based on the
order parameter based theory of phase transitions pioneered by
Landau. Specifically for a classical finite temperature phase
transition between two distinct ordered phases with different
broken symmetry characterized by two distinct order parameters,
Landau's theory forbids it from being generically second order. In
applying this intuition to the present phase transition we must
keep in mind two distinctions with the situation traditionally
considered in Landau theory. First the transition is a {\em
quantum} phase transition occurring at zero temperature. Second
the Kondo order in the heavy Fermi liquid phase is not captured by
any local order parameter.

Recent theoretical work\cite{deccp} has shown that magnetic quantum phase
transitions in two dimensional insulators can violate this natural
expectation from Landau theory. Specifically a direct second order
quantum phase transition between a Neel state (which breaks spin
rotation symmetry) and a spin-Peierls or valence bond solid state
(which breaks lattice rotation symmetry) was described for
spin-$1/2$ magnets on a square lattice. The resulting critical
theory is rather non-trivial, and is most naturally described in
terms of fractional spin (spinon) variables that interact with a
fluctuating emergent gauge field. Neither the spinons nor the
gauge field have any legitimacy in the excitation structure of the
two phases far away from the critical point. They become useful
degrees of freedom only right in the vicinity of the quantum
critical point. The analogy between these results and the heavy
fermi liquid critical points has been well explained in the
literature\cite{deccp,svsrev}. Two important features are worth emphasizing. 
First in these insulating magnets two distinct diverging length/time scales do appear for the
two fluctuating orders. Second the deconfined quantum critical point is the moral equivalent of a
non-Fermi liquid in the context of insulating magnets. The magnon spectral function measured in
neutron scattering experiments will be anomalously broad right at
the quantum critical point. This strong breakdown of the magnon as
a quasiparticle can be roughly understood as being due to decay
into two spinons at the critical point.

Since the original proposal of deconfined criticality a number of
numerical model calculations have sought to confirm the predicted
phenomena. Unfortunately the majority of such calculations seem to
find first order transitions or indications for unusual second
order transitions not predicted by the theory\cite{dccpnum}. A notable exception
may be the very recent calculations of Sandvik\cite{sdvkjq} on a fully $SU(2)$
symmetric spin-$1/2$ model on a square lattice where results
consistent with the theory of Ref. \cite{deccp} is obtained. Finding clear
numerical support for the theory of Ref. \cite{deccp} is clearly an
important open problem.

As far as lessons for the heavy fermion critical points go, it is
sufficient to consider the theoretical example of the same
transition in general SU(N) symmetric spin models\cite{RSSuN} and study it in
a systematic large-$N$ expansion. This provides an analytically
controlled example\cite{deccp} of such a deconfined quantum critical point and
therefore gives some legitimacy to exploring such phenomena in the heavy fermion context.

\subsection{`Local' quantum criticality}
A popular approach to the non-Fermi liquid heavy electron critical points is 
through the notion of `local' quantum criticality advocated by Si and coworkers\cite{qimiao} 
In this subsection we make several comments on this approach. Most importantly we 
point out that even within this framework the two-scale hypothesis emerges as a
natural possibility. 

The first comment is that the term `local' criticality does not imply that there is 
no diverging length scale. Rather the hypothesis is that the spatial correlation length 
for spin fluctuations near the ordering wavevector shows a simple mean field divergence with no anamolous exponent. However it is also hypothesized that there is a diverging correlation time
that is strongly not mean field like, and has a non-trivial anamolous exponent. Equivalently
right at criticality the spin susceptibility is assumed to take the form
\begin{equation}
\chi(\vec q,\omega) \sim \frac{1}{|\vec q - \vec Q|^2 + A(-i\omega)^{\alpha}}
\end{equation}
for $\vec q$ close to the ordering wavevector $\vec Q$; $A$ is a non-universal constant and $\alpha$ a universal exponent. This form is motivated by fits to neutron scattering data\cite{schroder} on $CeCu_{6-x}Au_x$. 
Existing theoretical justifications of this assumed form 
are based on Extended Dynamical Mean Field Theory(EDMFT) 
treatments of Kondo-Heisenberg models. It should be emphasized however that this kind of `locality' is to some extent built into EDMFT as it does not allow for $q$ dependence in the self
energies. 

The second comment is that it is important to distinguish the hypothesis of `local' criticality 
from the hypothesis that the non-fermi liquid behavior is associated with the transition from the heavy fermi liquid to the LMM state. The former is a specific statement about the 
structure of the singularities in, say the spin susceptibility at the critical point. The latter is a statement on the nature of the two phases involved in the transition. It is this latter 
hypothesis that requires large reconstruction of the Fermi surface and other related phenomena at the transition. It is by no means obvious that such a phase transition necessarily has the 
specific structure of critical singularities assumed in the `local' criticality scenario. Thus
experiments such as that in Ref. \cite{Silke} showing a jump in the Hall coefficient must be taken as evidence for the latter hypothesis rather than directly for the `local' criticality scenario. 

Finally we comment on the two-scale hypothesis within the framework of the EDMFT approach to the Kondo-Heisenberg lattice. In the presence of full $SU(2)$ spin symmetry, the EDMFT
calculations describe a transition between a Fermi liquid phase and a paramagnetic non-fermi liquid phase. The paramagnetic non-fermi liquid fixed point occurs at zero Kondo coupling. It thus seems correct to interpret this paramagnetic fixed point as a non-trivial gapless spin liquid formed from the local moments that is decoupled from conduction electrons. It is therefore to be seen as some version of an FL$^*$ phase. To argue that this transition describes the true transition from the heavy Fermi liquid to the LMM it is necessary to make two assumptions: first that the paramagnetic non-fermi liquid fixed point has a relevant perturbation that drives an instability toward a magnetically ordered phase (to be interpreted as LMM) and second that this perturbation is also irrelevant at the critical fixed point associated with the transition to the fermi liquid. In other words the magnetic ordering is driven by a perturbation that is dangerously irrelevant at the critical fixed point. It then follows that this transition will also be characterized by two diverging time scales. The time scale for magnetic ordering will diverge much faster than the time scale asociated with loss of Kondo screening exactly as proposed in Refs. \cite{svsrev,ssvprb}.

\subsection{Slave particle gauge theory}
If the two scale hypothesis is correct near the non-fermi liquid
heavy fermion critical point then it should be possible to
consider the Kondo and magnetic ordering phenomena as separate
crossovers. Consider first the Kondo-driven case of Section \ref{kd}. In
this case the primary transition involves loss of Kondo screening
and the subsequent jump in the Fermi volume. A theory for such a
transition is conveniently developed\cite{ssvprb} using a fermionic
representation of the local moment. (For prior calculations on an analogous transition in a
random Kondo lattice see Ref. \cite{georges}). At a mean field level the Kondo process is
described in terms of a hybridization amplitude between the $c$
and $f$ bands. The Heisenberg exchange leads to the presence of
some dispersion in the $f$-band. As a function of the ratio
$J_K/J_H$ two distinct states are found. For large Kondo coupling
the Kondo hybridization is non-zero and a heavy Fermi liquid state
obtains. For small Kondo coupling on the other hand the
hybridization is zero. The $c$-electrons are then decoupled from
the local moments which have settled into a spin liquid state.
This is an example of an FL$^*$ phase. The transition between the
two phases is second order in mean field theory, and is
accompanied by a jump in the Fermi volume. The critical point
describes a non-Fermi liquid state.

This calculation provides an explicit realization of the scenario
(see the end of Section \ref{slfs}) in which the Fermi surface of the
heavy Fermi liquid consists of two sheets near the transition. All
the important action is associated with just one sheet (the hot
sheet) while the other cold sheet remains more or less a benign
spectator. In particular the quasiparticle residue vanishes all
over the hot sheet but stays finite on the cold sheet.

Fluctuation effects beyond the mean field theory were also
examined in Ref. \cite{ssvprb}, and shown to lead to singular non-Fermi
liquid temperature dependences for a number of physical quantities
(for example specific heat and resistivity). Very recently Coleman
et al\cite{cmsprb} have studied the evolution of the zero temperature transport
properties across the transition, and argued that both the
longitudinal and Hall resistivities will jump at the critical
point.

It remains to be seen whether this kind of transition (to a
suitable unstable FL$^*$ phase) can provide a platform for
describing a Kondo-driven transition between the heavy Fermi
liquid and the LMM state.

It is natural to expect that the magnetism-driven scenario of
Section \ref{md} is better accessed by employing a description of the
spins in terms of Schwinger bosons advocated in Ref. \cite{pepin}. 
It again remains to be seen whether such an approach can
eventually lead to a theory of the LMM- heavy fermi liquid
transition.

\section{Conclusion}
\label{concl}

In this paper we have discussed some ideas on the non-fermi liquid behavior found near the magnetic quantum critical point in some heavy fermion materials. Clearly there is much work 
left (and more ideas needed) to understand the basic physics of this non-fermi liquid.  Our goal was to outline in a general way some of the theoretical issues and to 
motivate experiments. Here we reiterate some of the most important questions that 
future experiments will hopefully clarify. 

(i) It is important to establish a clear connection (if any) between the nature of the magnetic phase to which the transition happens (LMM or SDW) and the occurence of non-fermi liquid criticality. If as in the simple situation discussed in Section \ref{mmkl} the two magnetic metals are 
sharply distinct then it may be possible to establish the nature of the magnetic phase through measurements far from the critical point. 

(ii) What is the fate of the Fermi surface right at the quantum critical ground state? As discussed
in Section \ref{slfs} it seems reasonable to assume that a sharp Fermi surface albeit with power law quasiparticles survives. If so will it be super-large or will it have several sheets with some sheets with power law quasiparticles and the others with ordinary (delta function) quasiparticles? 

(iii) Are the fermi surface reconstruction and magnetic ordering phenomena characterized by two 
distinct diverging length/time scales? As we have discussed such a possibility is theoretically appealing and finds support in other analogous simpler theoretical problems.

I thank S. Sachdev, Q. Si, and M. Vojta for several useful discussions. This work was supported by
funding from the NEC Corporation, the Alfred P. Sloan
Foundation, and an award from the The Research Corporation.



\end{document}